 \documentclass[prd,preprintnumbers,amsmath,amssymb,twocolumn]{revtex4-2} 

\usepackage{amsmath} 
\usepackage{amsfonts} 
\usepackage{graphicx}
\usepackage{dcolumn}
\usepackage{bm}
\usepackage{xcolor}
\usepackage{hyperref}

\begin{document}

\title{Learning dynamics on the picosecond timescale in a superconducting synapse structure}

\author{Ken Segall, Leon Nichols, Will Friend, and Steven B. Kaplan}

\affiliation{Department of Physics and Astronomy, Colgate University}
\begin{abstract}

 Conventional Artificial Intelligence (AI) systems are running into limitations in terms of training time and energy.  Following the principles of the human brain, spiking neural networks trained with unsupervised learning offer a faster, more energy-efficient alternative.  However, the dynamics of spiking, learning, and forgetting become more complicated in such schemes.  Here we study a superconducting electronics implementation of a learning synapse and experimentally measure its spiking dynamics.  By pulsing the system with a superconducting neuron, we show that a superconducting inductor can dynamically hold the synaptic weight with updates due to learning and forgetting.  Learning can be stopped by slowing down the arrival time of the post-synaptic pulse, in accordance with the Spike-Timing Dependent Plasticity paradigm.  We find excellent agreement with circuit simulations, and by fitting the turn-on of the pulsing frequency, we confirm a learning time of 16.1 ± 1 ps.  The power dissipation in the learning part of the synapse is less than one attojoule per learning event.  This leads to the possibility of an extremely fast and energy-efficient learning processor.

\end{abstract}


\maketitle 

\section{Introduction}\label{sec:Intro}
Over the past two decades, the human brain has become a powerful inspiration for computing systems.\cite{furber_neural_2006}  Artificial neural networks and deep learning\cite{lecun_deep_2015}, which are the powerhouses of today’s AI (Artificial Intelligence) systems, are based on the idea of synaptic weighting, a fundamental principle in how neurons connect to each other in the brain.  Temporal spiking, which has been the inspiration for spiking neural networks (SNNs) and neuromorphic computing\cite{kudithipudi_neuromorphic_2025,frenkel_bottom-up_2023}, is based on the dynamics of action potentials of neurons in the brain.  Given that the brain is fault-tolerant, parallel, adaptable, and highly energy efficient, continuing to mimic its operation is a worthwhile strategy.

Looking for further inspiration, another important aspect of the brain’s operation is learning, in which synaptic connections are strengthened and weakened over time due to the firing patterns of the adjoining neurons.  The brain utilizes fully unsupervised learning, in stark contrast to today’s AI systems, which are mostly trained offline with supervised learning utilizing back propagation algorithms.  The time, energy, and monetary cost of supervised learning has been well-documented\cite{sevilla_compute_2022}, so the incorporation of unsupervised learning into AI systems is certainly a promising avenue forward.  However, unsupervised learning in a spiking system results in far more complicated dynamics as well as challenges in training. 

Superconducting electronics offers a promising platform for neuromorphic computing.\cite{schneider_supermind_2022,feldhoff_short-_2024,chalkiadakis_dynamical_2022,goteti_collective_2024,goteti_superconducting_2022,hirose_pulsed_2007,jardine_hybrid_2023,onomi_improved_2014,primavera_programmable_2024,razmkhah_hybrid_2024,schegolev_bio-inspired_2023,schneider_self-training_2025,schneider_ultralow_2018,shainline_optoelectronic_2021,shainline_superconducting_2019,semenov_biosfq_2023,yamanashi_pseudo_2013,cheng_toward_2021,toomey_design_2019,toomey_superconducting_2020}  Spiking and thresholding operations are inherent to the physics of Josephson junctions, and superconducting transmission lines can carry action potential pulses over long distances without distortion.  Synaptic weighting is possible with inductive division and transformer coupling.  The Josephson junction (JJ) neuron, developed in our research group\cite{crotty_josephson_2010}, is highly biologically realistic, demonstrating 19 of the 20 possible Izhikevich behaviors \cite{crotty_biologically_2023}, and has been experimentally shown to exhibit some collective neural behavior like phase-flip bifurcations.\cite{segall_synchronization_2017}  System projections for a neural network based on superconducting electronics show at least an order of magnitude improvement in energy efficiency and speed compared to semiconductor platforms.\cite{tschirhart_brainfreeze_2021} 

In this work, we experimentally study the learning dynamics of a superconducting synapse structure coupled with a Josephson junction neuron.  We employ a learning gate and memory structure, which was described in prior work.\cite{segall_superconducting_2023}  We show that a superconducting inductor can dynamically hold the value of a synaptic weight with updates due to learning and forgetting.  A weakly-coupled Superconducting Quantum Interference Device (SQUID) can read the synaptic weight with good fidelity.  The rate of learning events depends on both the firing frequency of the JJ neuron and the arrival time delay between the two pulses in the synapses, in accordance with the spike-timing dependent plasticity (STDP) paradigm. \cite{dan_spike_2004,kandel_principles_2000}  The fundamental learning time can be estimated by slowing down one pulse relative to the other.  We demonstrate that the dynamics of the synaptic weight are well-explained by the processes of learning and forgetting.  Circuit simulations agree very well with experiment, and by fitting the turn-on of learning events with current, we confirm a learning time of 16.1 ± 1 ps with a power dissipation of less than one attojoule.  This points toward the possibility of a fast and energy-efficient learning processor working in a manner similar to the human brain.

\section{Methods}\label{sec:Methods}
\begin{figure*}[t!]
    \centering
    \includegraphics[width=16cm, height=8cm]{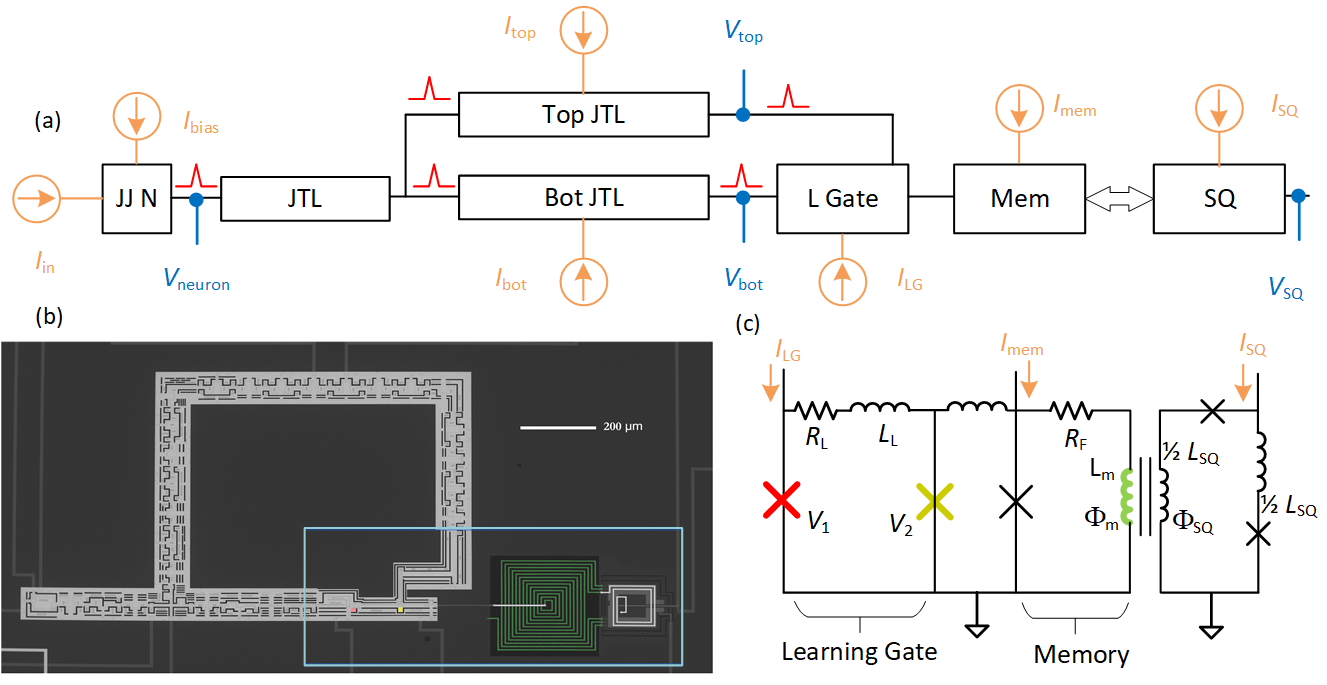}
    \caption{Circuit schematic and image.  (a) Circuit schematic of the whole circuit, showing the JJ Neuron (JJ N), top and bottom JTLs, learning gate (L Gate), memory inductor (Mem) and SQUID (SQ).  (b) SEM micrograph of the circuit.  The memory inductor is artificially colored green and the two junctions of the learning gate are colored red and yellow.  The light blue box indicates the learning part of the circuit.  (c) Circuit diagram of the learning gate, memory and SQUID with the electrical parameters labeled. }
    \label{fig:cirDiagram}
\end{figure*}
\subsection{Circuit Description}
The circuit schematic is shown in Fig. 1a.  A single JJ neuron is used to pulse the system.  Its output is coupled to a Josephson Transmission Line (JTL), which is split into two branches, referred to as the bottom and  top JTLs.  The pulses traveling along the two branches are designed to mimic a pre-synaptic pulse (bottom) and post-synaptic pulse (top); their transit times can be adjusted by separate bias currents.  The two pulses are recombined at the learning gate, which outputs a pulse if the two pulses arrive within a time window $\tau_\mathrm{L}$, called the learning time.  These learning pulses are coupled to a memory loop, which contains the memory inductor $L_\mathrm{m}$ and the forgetting resistor $R_\mathrm{F}$.  The memory inductor is weakly coupled to a SQUID to measure its flux.  Fig. 1b shows a Scanning Electron Microscope (SEM) image of the circuit.  Fig. 1c shows the circuit schematic of the learning gate, memory cell, and SQUID, and labels some important electrical quantities.

The circuit requires seven currents for bias and control.  The firing of the JJ neuron is set by two currents, labeled as the bias current $I_\mathrm{bias}$ and the input current $I_\mathrm{in}$; these labels are consistent with previous work.  The bottom and top JTL are biased with currents $I_\mathrm{bot}$ and $I_\mathrm{top}$, respectively.  The learning gate, the memory cell and the SQUID are biased with currents $I_\mathrm{LG}$, $I_\mathrm{mem}$, and $I_{SQ}$.  

After these seven currents are set, four different voltages are measured.  The pulsing of the neuron and the two transmission lines are indicated by their voltages $V_\mathrm{neuron}$, $V_\mathrm{bot}$, and $V_\mathrm{top}$; the SQUID voltage $V_\mathrm{SQ}$ is proportional to the flux coupled into the SQUID from the memory inductor.  Our voltage measurements are on the millisecond timescale, whereas the spiking times are on the picosecond timescale; thus our measurements of voltage represent the long-term, DC (Direct Current) average.  In this limit the DC voltages are proportional to the junction spiking frequency via the Josephson relation, 2.07 $\mu \mathrm{V/GHz}$, and so we can opt to display our measured voltages as spiking frequencies, which we do for most of the figures in the results section.  

The circuit was fabricated at Hypres Inc. in a niobium trilayer process with a current density of $J_c$ = 1~kA/$cm^2$.  The junctions yielded well and showed good current-voltage curves.  Independent measurements on three different control structures located on the chip found the critical currents to be slightly higher than their design values while the inductances and resistances were slighltly lower.  These were accounted for in the circuit simulations (see below). 

\subsection{Experiments}
Experiments were performed between 2.8 and 4.2 K in a cryogen-free refrigerator from FormFactor.  The bias lines were heavily filtered with low-pass filters and Echosorb filters\cite{santavicca_impedance-matched_2008}, and all measurement electronics were battery powered.  The input current $I_\mathrm{in}$ was sourced by an AC function generator coupled through a transformer to a balanced bias circuit, while the remaining currents were sourced by custom-made, single-ended current supplies made from precision voltage regulators or programmable voltage supplies.  DC voltages were measured by Analog Devices instrumentation amplifiers.  Typical sweeps involved choosing two currents to vary, such as $I_\mathrm{in}$ and $I_\mathrm{bias}$, while keeping the other currents constant.  After data collection, several of the voltages were plotted versus those two currents in a 2-D color plot (see supplemental section).

\begin{figure*}[t!]
    \centering
    \includegraphics[width=16cm, height=6cm]{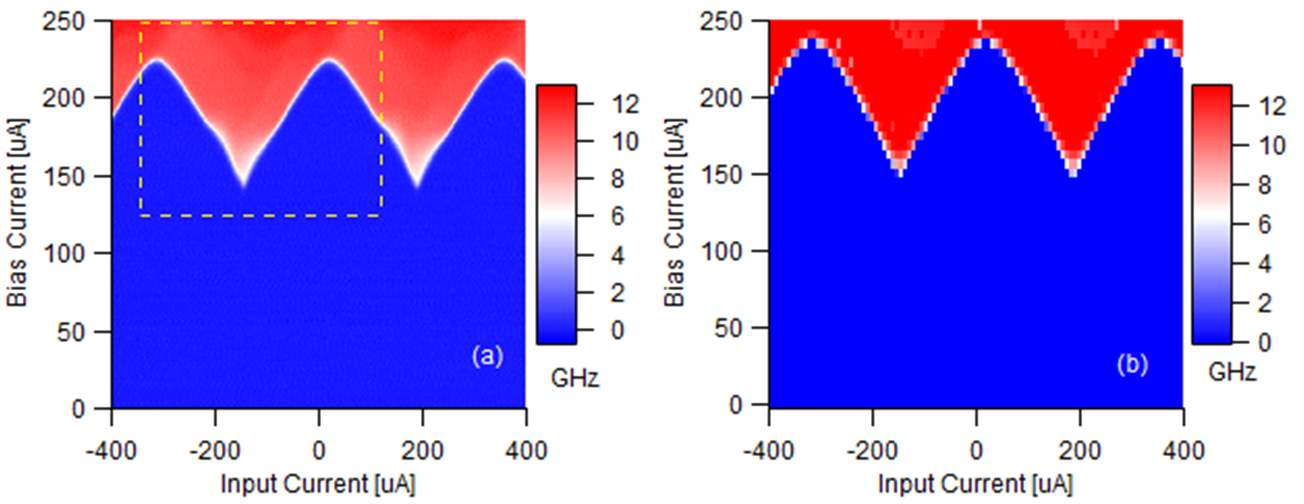}
    \caption{Neuron frequency as a function of input and bias current.  (a) Experiment.  The yellow rectangle indicates a subsection for comparison in Fig. 4. (b) Simulation with the same parameters. }
    \label{fig:pulseFreq}
\end{figure*}

\subsection{Simulations}
Simulations were performed with WRSPICE.  After setting values of the seven currents, transient analyses with timesteps of 0.02 ps were performed for periods of 15 to 25 ns, long enough for the system to reach a steady state.  The average voltages were then recorded.  Currents and voltages were swept and plotted in the same way as the experimental data.  Circuit parameters including resistances, critical currents and inductances in the WRSPICE file were set to the design values with three adjustable, chip-wide scale parameters: the overall current density ($J_c$), the sheet resistance of the resistor layer ($R_\square$), and a scale factor for the inductance values ($L_\mathrm{factor}$).  We constrained these three parameters to be within one standard deviation of the independent control measurements and then chose the values to best fit the data.  The values used in the simulation were $J_c$ = $1.08~kA/cm^2$ (compared to $1.00~kA/cm^2$), $R_\square = 3.6~\Omega /\square$ (compared to $4.0~\Omega /\square$), and $L_\mathrm{factor}$ = 0.79 (compared to 1.0).

\section{Results}\label{sec:Results}

\subsection{Pulsing Frequency of Neuron}
The overall approach of our measurements was to pulse the JJ neuron at different frequencies by varying its currents, record the voltages in the circuit, and then explain those voltages with the dynamical effects of learning.  The JJ neuron is driven by two currents, the input current ($I_\mathrm{in}$) and the bias current ($I_\mathrm{bias}$), and changing those two currents varies the neuron frequency from about 0 to 13 GHz in a highly non-trivial way due to the nonlinearity of the JJ neuron.  If the JTLs are biased properly and there are learning events at the learning gate, then the whole system will pulse at the same frequency as the JJ neuron.  We measure this frequency with the voltage across the JJ neuron ($V_\mathrm{neuron}$), the voltage at the top JTL ($V_\mathrm{top}$), and the voltage at the bottom JTL ($V_\mathrm{bot}$).  After converting those voltages to a spking frequency, we find that under proper steady-state biasing that all three of these frequencies are the same to within ±0.2 GHz for all values of $I_\mathrm{in}$ and $I_\mathrm{bias}$, with no systematic trend in the differences (see supplemental section).  Arbitrarily choosing $V_\mathrm{top}$ to display, Fig. 2a shows a color plot of $V_\mathrm{top}$ as a function of the input current (horizontal axis) and bias current (vertical axis).  The SQUID-like modulation of the JJ neuron's frequency is evident.  This measured color plot contains a lot of information, 250 x 1000 points (bias current x input current).  Similar plots were used in previous work by our group.\cite{segall_synchronization_2017}

Fig. 2b shows a calculation of the pulsing frequency measured in Fig. 2a using WRSPICE.  Good agreement is seen, except for some slight deviations near the minimum of the SQUID-like curve where the neuron loop is maximally frustrated, at the point where the induced flux in the JJ neuron loop is about ($\Phi_0$/2).  The plot in Fig. 2b is lower resolution (42 x 100) than the measurement due to computational time limitations; the simulation time per point is about 15,000 times longer than the measurement time, an effect also documented in previous work.\cite{segall_synchronization_2017}

\begin{figure}[t!]
    \centering
    \includegraphics[width=8cm, height=6cm]{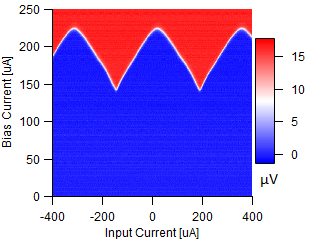}
    \caption{: Voltage in the SQUID as a function of $I_\mathrm{bias}$ and $I_\mathrm{in}$ measured simultaneously with the pulsing frequency in Figure 2a, over the same range of points. }
    \label{fig:squidV}
\end{figure}

\subsection{Learning Effects in the SQUID signal}
\begin{figure*}[t!]
    \centering
    \includegraphics[width=16cm, height=11cm]{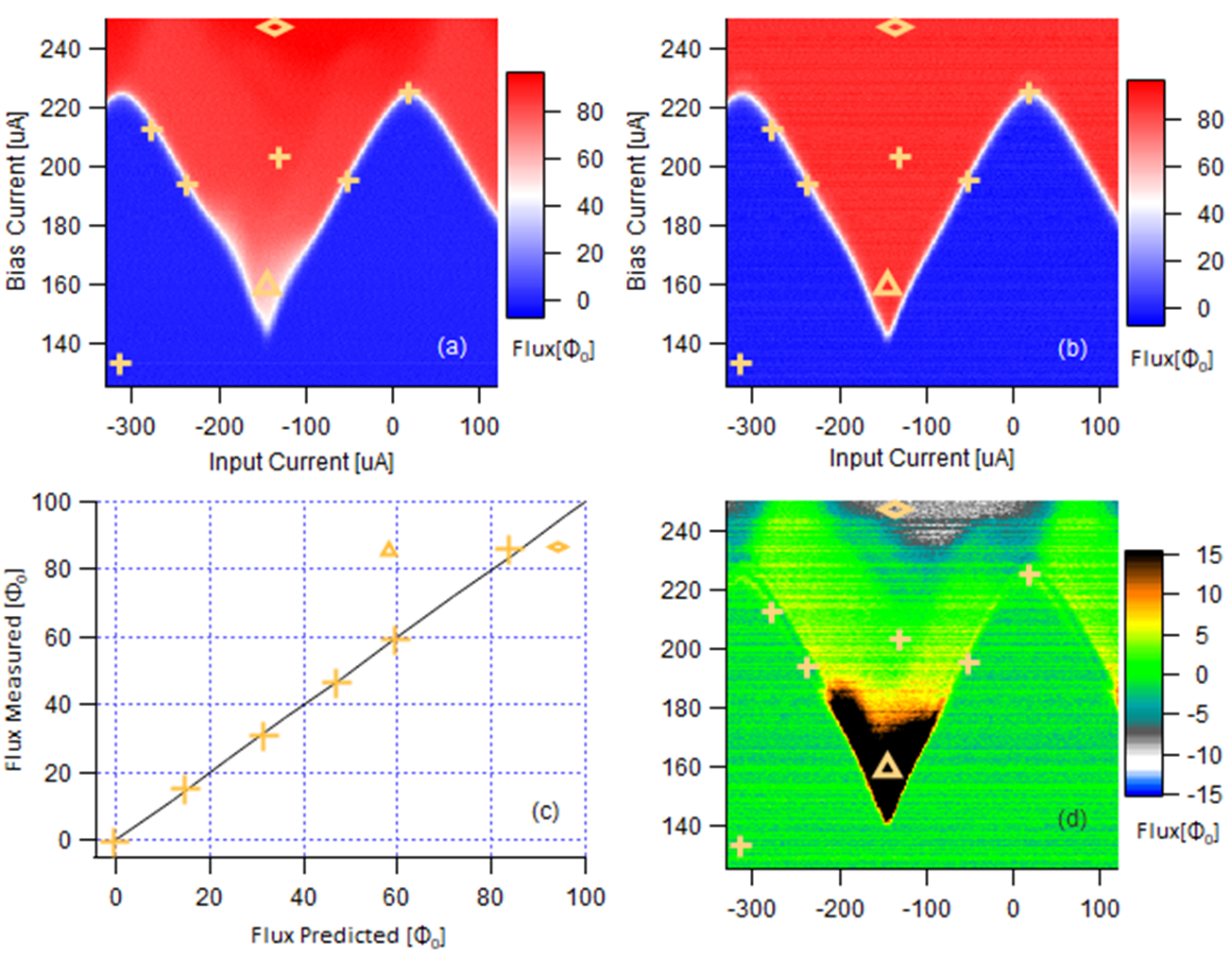}
    \caption{Comparison between predicted and measured value of the flux $\overline{\Phi}_\mathrm{m}$ in the memory loop in units of the flux quantum.  (a) Predicted value of $\overline{\Phi}_\mathrm{m}$.  The range is over the set of points indicated by the yellow rectangle in Fig. 2a.  (b) Measured value of $\overline{\Phi}_\mathrm{m}$ over the same range.  (c) Measured versus predicted flux for the small set of points indicated by the symbols indicated in (a) and (b).  (d) Difference between measured and predicted flux, in the same range as (a) and (b).  }
    \label{fig:compFig}
\end{figure*}
The synaptic weight of the synapse is held by the flux $\Phi_\mathrm{m}$ in the memory inductor $L_\mathrm{m}$.  To measure $\Phi_\mathrm{m}$ directly, we measure the voltage across the SQUID.  Fig. 3 shows the voltage across the SQUID in $\mu$V measured simultaneously with the pulsing frequency in Fig. 2a.  The similarity in the color plots indicates that when the neuron is pulsing fast, the signal in the SQUID is high, and when the pulsing is slow, the signal is low.

To better understand the relationship between the pulsing of the neuron (Fig. 2a) and the signal in the SQUID (Fig. 3), we derive an expression for the memory flux $\Phi_\mathrm{m}$ in terms of the pulsing frequency of the neuron.  Initially $\Phi_\mathrm{m}$ is zero, when the neuron first starts pulsing, but very quickly increases to a steady-state value, which we denote by $\overline{\Phi}_\mathrm{m}$.  This value is set by the competition between learning, which puts flux into the memory loop, and forgetting, which removes flux from the loop.  Learning events, each of which put a flux $\Phi_0$ into the loop, occur at a rate of $1/\tau_E=V_m/\Phi_0$, where $V_m$ is the voltage across the memory junction.  Since each pulse of the memory junction results from a pulse of the JJ neuron and the JTL junctions, we can substitute $V_\mathrm{top}$ for $V_m$.  Meanwhile, forgetting occurs continuously at a rate of $1/\tau_F=R_\mathrm{F}/L_m$, due to the decay of circulating current in the memory loop.  The differential equation for $\Phi_\mathrm{m}$ is then:
\begin{equation}
\frac{d\Phi_m}{dt}=+\frac{\Phi_0}{\tau_E}-\frac{\Phi_m}{\tau_F},
\label{fluxDiffEq}
\end{equation}
which has a steady-state solution of:
\begin{equation}
\overline{\Phi}_m=\frac{L_m}{R_F}V_{top}.
\label{fluxSoln}
\end{equation}
All measured voltages are in the long-time average, much longer than the time for $\Phi_\mathrm{m}$ to reach steady-state, so there is no need to explicitly average the right-hand side of eq. (2); similarly for eq. (5) below. Meanwhile, the SQUID is mutually coupled to $L_\mathrm{m}$ through the mutual inductance $M$.  The flux in the SQUID $\Phi_{SQ}$ is related to $\Phi_\mathrm{m}$ by:
\begin{equation}
\Phi_{SQ}=\frac{M}{L_m}\Phi_m,
\label{SQflux}
\end{equation}
and the voltage change across the SQUID is then given by:
\begin{equation}
V_{SQ}=\frac{R_{SQ}}{L_{SQ}}\Phi_{SQ}.
\label{SQVolt}
\end{equation}
Equation (4) assumes optimal biasing and a SQUID flux much less than $\Phi_0$; this is discussed more below.  Combining (3) and (4) and taking the steady-state we find:
\begin{equation}
\overline{\Phi}_m=\frac{L_mL_{SQ}}{MR_{SQ}}V_{SQ}.
\label{SQVolt}
\end{equation}
Equation (2) gives a predicted value of the flux in the memory loop due to the pulsing of the neuron; equation (5) gives the measured value of flux in the memory loop from the voltage in the SQUID.  The comparison between the two is shown in Fig. 4.  Fig. 4a shows the predicted flux using eq. (2) and the data from Fig. 2a for $V_\mathrm{top}$, while Fig. 4b uses the measured flux using eq. (5) and the data from Fig. 3 for the signal on the SQUID.  Figs. 4a and 4b use the set of points inside the yellow rectangle in Fig. 2a, to show more detail.  The color scale is in units of the flux quantum $\Phi_0$. 

The agreement between the predicted and measured values of $\overline{\Phi}_\mathrm{m}$ is excellent for the majority of the 250,000 points measured.  Note that there are essentially no adjustable parameters in equations (2) and (5), other than the adjustments made to the chip-wide scale factors for inductance and resistance.  Fig. 4c shows a plot of the measured versus predicted flux $\overline{\Phi}_\mathrm{m}$ for the small set of points indicated in Figs. 4a and 4b.  The crosses show places in the color plot where the predicted and measured values agree with each other, while the diamond indicates a region where the measured value is smaller than the predicted value, and the triangle indicates a region where the measured value is larger than predicted value; these disagreements are discussed below.  Fig. 4d shows a color plot of the difference between measured and predicted (measured minus predicted).  The majority of the plot is green, where there is good agreement between measured and predicted.

The two places of disagreement in Fig. 4d are (1) at high values of $I_\mathrm{bias}$, where the measured values are a little less than the predicted values, appearing as grayish, and (2) near the minima of the SQUID-like curve, where the measured values are higher than predicted, appearing as dark brown or black.  The disagreement at high values of $I_\mathrm{bias}$ occurs because as the flux $\Phi_\mathrm{m}$ gets large, the small-signal limit starts to be exceeded and the gain of the SQUID drops.  The disagreement near the minima of curves is unexplained at this point; it could perhaps be due to some hysteresis or bi-stability in the maximally frustrated region.  Interestingly, this disagreement occurs in a region similar to that between the measured and simulated pulsing frequency in Fig. 2. This will be a subject of future work. 

The results of this section show that we can understand and measure the dynamical effects of learning.  In previous work we showed more detailed time-dependent simulations of this synaptic weight changing over different timescales under different learning conditions.\cite{segall_superconducting_2023}

\subsection{Learning Time Estimation}

\begin{figure}[t!]
    \centering
    \includegraphics[width=8cm, height=6cm]{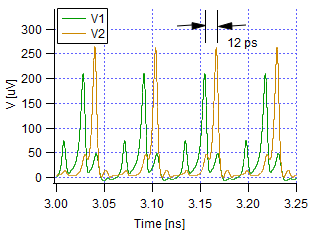}
    \caption{Simulation of the dynamics of the learning gate.  The voltages of the two junctions, $V_1$ and $V_2$, are plotted versus time, starting 3~ns into the simulation.  The time between the arrival of the two pulses is about 12~ps. }
    \label{fig:LGdynamics}
\end{figure}

The previous section highlighted the tradeoff of learning and forgetting, and the successful fitting of the data can be seen as confirming the forgetting time $\tau_F = L_m/R_\mathrm{F} = 7.8$ ns.  The learning time, which is the difference between the arrival of the pre-synaptic and post-synaptic pulses at the learning gate, is much shorter: $\tau_L = L_\mathrm{L}/R_\mathrm{L} = 16.1$ ps.  In this section we use measurements and simulations to confirm that the learning time is indeed that short.

\begin{figure*}[t!]
    \centering
    \includegraphics[width=16cm, height=6cm]{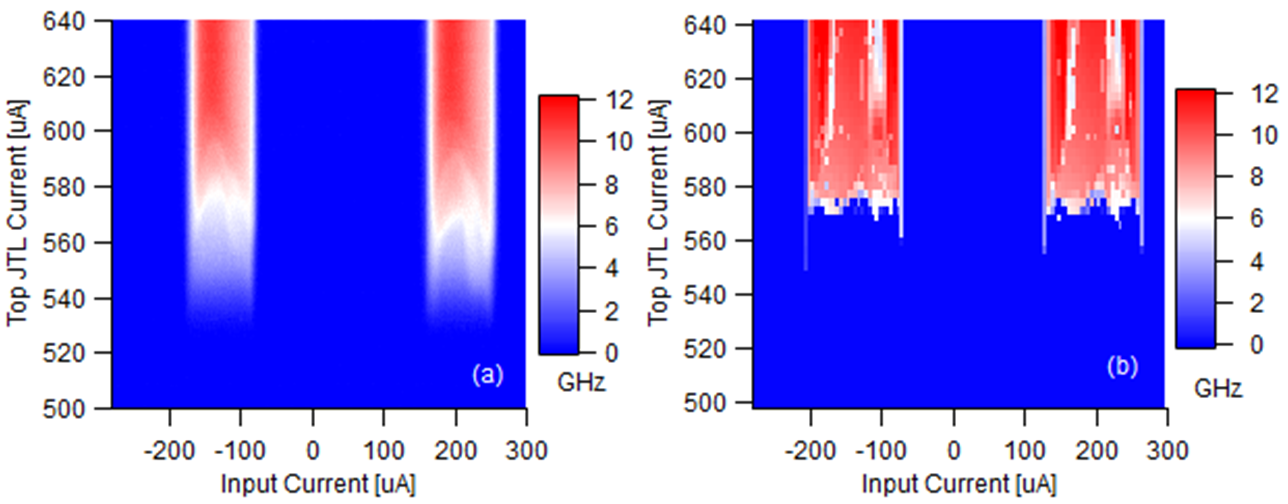}
    \caption{Spiking frequency of the top JTL as a function of input current and top JTL current, for a fixed neuron bias current of 190 $\mu A$.  (a) Experiment, and (b) Simulation.}
    \label{fig:TopJTLcolor}
\end{figure*}

Fig. 5 shows a simulation with $I_\mathrm{bias} = 200 \mu A$ and $I_\mathrm{in} = -150 \mu A$, in the region where there is learning on every pulse.  We plot the voltage across the two junctions in the learning gate, $V_1$ and $V_2$, for a time starting 3 ns into the simulation, after all of the transients have died out.  The time between consecutive peaks of $V_1$ and $V_2$ is about 12 ps, within the expected learning time of 16.1 ps, and thus consistent with the STDP picture.  Unfortunately, we are not able to observe these pulses directly, as they are too fast to couple out of the cryostat and digitize.  Arguably, by finding simulation parameters that give a good match with the experiment, like in Fig. 2a, combined with the prediction of those simulations, like in Fig. 5, we have indeed confirmed learning on a timescale that is on that order.  However, we can go a step further and use another feature of the data that is sensitive to the learning time, described below.      

The measurements in the previous section were taken in a regime where learning is occurring on every pulse, namely every pulse of the neuron results in a pulse of the memory junction.   This is because, for those data, the top and bottom JTLs were biased in such a way as to allow proper pulse propagation.  As shown in Fig. 5, under those conditions the pulses from the top and bottom JTLs arrive at the learning gate within the learning time $\tau_L$ of each other.  We can violate this condition, however, by decreasing the current on the top JTL.  This slows down the top pulse with respect to the bottom pulse and prevents it from arriving at the learning gate within the learning time of the bottom pulse.  At some point all pulsing stops, since the boundary conditions of the JTLs are no longer favorable for nonlinear wave propagation of single flux quantum pulses.

A way to visualize this is experimentally is to measure the spiking frequency in a similar way to Fig. 2a, but instead vary the top JTL current while keeping the neuron bias fixed.  Fig. 6a shows a color plot of the pulsing frequency along the top JTL as a function of neuron input current (horizontal axis) and top JTL bias current (vertical axis) at a fixed bias current of 190~$\mu A$.  If we move horizontally in Fig. 2a at a fixed bias current of around 190 $\mu A$, we expect a region of firing at an input current of about -150 $\mu A$ and again around +200 $\mu A$; these are indeed seen in Fig. 6a, appearing as column features in the plot.  Fig. 6b shows the equivalent WRSPICE simulation and the same features are seen as well.

\begin{figure}[t!]
    \centering
    \includegraphics[width=8cm, height=6cm]{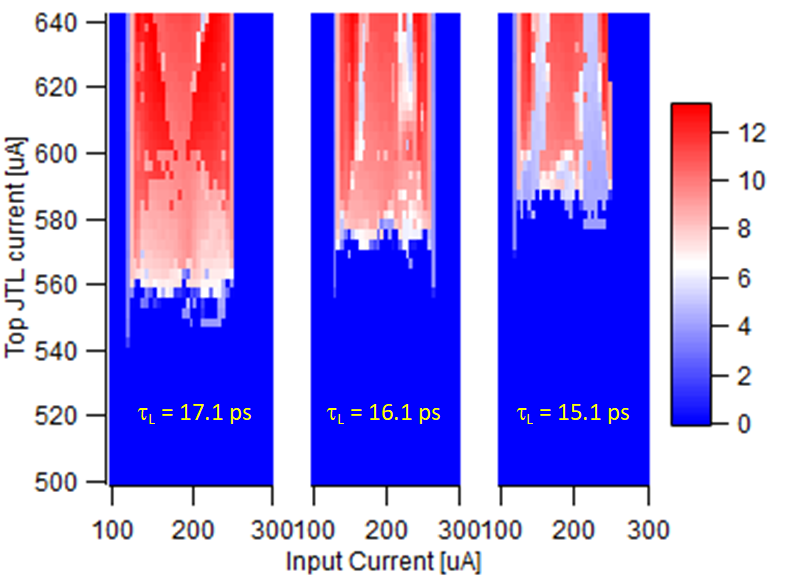}
    \caption{Spiking frequency of the top JTL as a function of neuron input current and top JTL current, for three different values of the learning time.}
    \label{fig:LearningTimeFit}
\end{figure}

The column features disappear as the top JTL current is reduced below a certain value, around 540 $\mu A$ or so in Fig. 6a.  We then associate this point with the violation of the learning condition, since the pulse along the top branch has slowed enough to no longer be able to arrive at the learning gate in time.  Fig. 7 shows a simulation of the right column feature in Fig. 6b for three different values of the learning time $\tau_L$: 17.1 ps (left panel), 16.1 ps (middle panel), and 15.1 ps (right panel).  We can see the termination point of the column is extremely sensitive to the value of $\tau_L$.  Part of this sensitivity comes from resonant steps in the JTL and is discussed below.

In Fig. 8 we take an average of the pulsing frequency over neuron input currents from 110 $\mu A$ to 250 $\mu A$, which essentially reduces the right column feature in Fig. 6 to a single line, and plot that average versus Top JTL current.  We also show curves for the data and the three different simulated learning times from Fig. 7.  The wider turn-on region of the experiment is due to thermal activation, which is not included in the simulation.  Its width is about 2-3 percent of the critical current of the top JTL, which is consistent with thermal activation at the experimental temperature of 3 K.\cite{garg_escape-field_1995}  The learning time of 16.1 ps is the best match, but with other uncertainties in the fitting procedure we conservatively estimate $\tau_L$ = 16.1 ± 1 ps.

Figs. 6 and 7 show interesting structure in the color plots, with white and red sections alternating with each other in the pulsing regions.  These structures result from resonant steps in the top JTL, which have been well studied \cite{ustinov_experimental_1995,ustinov_fluxon_1993,watanabe_dynamics_1996,watanabe_whirling_1995} in the past.  Although our structure is a little more complicated than these studies, since there is an attached neuron of varying frequency and a boundary formed by the learning gate, low-voltage resonances are still expected in the 6-16 V range (3-8 GHz), resulting in steps of very small dynamic resistance $(dV/dI)$.  Below we show that the small value of $(dV/dI)$ is why the pulse turn-on in Figs. 7 and 8 is so sensitive to the learning time.

\begin{figure}[t!]
    \centering
    \includegraphics[width=8cm, height=6cm]{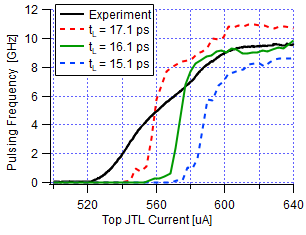}
    \caption{Spiking frequency as a function of top JTL current only.   The data are compared for three different values of the learning time. }
    \label{fig:squidV}
\end{figure}

Looking at Fig. 8, we can roughly estimate that the 2 ps change in learning time between the red and blue dotted curves is associated with about a 30 $\mu A$ change in the top JTL bias current, at a spiking frequency of about 5 GHz (voltage of about 10 $\mu V$).  In other words, if the top JTL current increases by 30 $\mu A$, pulses arrive at the learning gate 2 ps faster.  Approximating pulse propagation along the JTL as simple wave propagation, we can write that $d/t_\mathrm{travel} = \lambda /T$, where $\lambda$ is the wavelength, $T$ is the temporal spiking period of the junction, and $t_{travel}$ is the travel time.  The change in travel time is related to to a change in period by $\Delta t_\mathrm{travel} = (d/\lambda)\Delta T$.  Changes in the period are related to changes in current by:
\begin{equation}
\Delta T=(\frac{dT}{dI})\Delta I=(-\frac{\Phi_0}{V^2})(\frac{dV}{dI})\Delta I,
\label{DeltaT}
\end{equation}
so that with a small $(dV/dI)$ one needs a larger $\Delta I$ for the same $\Delta T$.  (Note that the minus sign simply indicates that decreasing the current increases the period.)  Using our numbers we estimate that $(dV/dI)$ is about 20~$m\Omega$ per junction, about two orders of magnitude smaller than the actual shunt resistance, confirming the presence of resonances.  This increased sensitivity is interesting and has yet to be observed or understood in a neuromorphic context.  It could be utilized to improve performance in future devices.

We can also use the information from the time-dependent part of the simulation, as in Fig. 5, to estimate the power dissipation in the learning part of the circuit.  We calculate the power as $V^2/R$ for each junction and find that the two junctions in the learning gate ($V_1$ and $V_2$) dissipate 0.117 and 0.104 attojoules per learning event, respectively, and the memory junction dissipates 0.259 attojoules per learning event, all for the biasing conditions of Fig. 5. The total is 0.48 attojoules, consistent with estimates in previous studies.\cite{tschirhart_brainfreeze_2021}    

\section{Conclusion}
We have measured a superconducting synapse structure that contains a superconducting inductor to hold the synaptic weight and a learning gate to implement a Spike Timing Dependent Plasticity scheme for unsupervised learning.  By pulsing the system with a Josephson junction neuron, we explore the dynamics of this synapse structure on the picosecond time scale.  The synaptic weight is monitored by a weakly-coupled SQUID and its signal is well-explained by the processes of learning and forgetting.  Learning events can be stopped by slowing down the arrival of the postsynaptic pulse.  Fitting the turn-on of learning events allows us to confirm a learning time of 16.1 ± 1 ps, with a power dissipation of less than half of an attojoule.  Circuit simulations agree extremely well with experimental results, showing that we understand the important processes at play.  The sensitivity of the system is enhanced by resonances in the Josephson transmission lines, an effect that may be useful for future devices.  The picosecond learning time and low energy dissipation point toward the possibility of an extremely fast, energy-efficient Spiking Neural Network made from superconducting electronics.

\section*{Acknowledgments}
Thanks to M.E. Parks and M.L. Schneider for manuscript review, D. Schult for useful discussions, and H. Benze for technical assistance.

\appendix

\bibliography{LDpaperLibrary}
\bibliographystyle{unsrt}

\end{document}